# Stark effect on a geometry defined by a cake' slice.

Jorge-Alejandro Reyes-Esqueda[a,*], Carlos I. Mendoza[b], Marcelo del Castillo-Mussot[a], Gerardo J. Vázquez[a].

[a]Instituto de Física, UNAM, 04510, P. O. Box 20-364, 01000, México, D. F., México.
[b]Instituto de Investigaciones en Materiales, UNAM, P. O. Box 70-360, 04510, México, D. F., México.

**Abstract**

By using a variational calculation, we study the effect of an external applied electric field on the ground state of electrons confined in a quantum box with a geometry defined by a slice of a cake. This geometry is a first approximation for a tip of a cantilever of an Atomic Force Microscope (AFM). By modeling the tip with the slice, we calculate the electronic ground state energy as function of the slice's diameter, its angular aperture, its thickness and the intensity of the external electric field applied along the slice. For the applied field pointing to the wider part of the slice, a confining electronic effect in the opposite side is clearly observed. This effect is sharper as the angular slice's aperture is smaller and there is more radial space to manifest itself.

---

[*] Corresponding author. Tel.: +52 55 5622-5184; Fax: +52 55 5616-1535.
*E-mail address*: reyes@fisica.unam.mx (J.-A. Reyes-Esqueda)

## 1. Introduction

In the last few years, there has been an increasing interest for studying the properties of nanostructured materials in looking for new technological applications. Specially, the behavior of their electronic and optical properties in the presence of an external electric field has received special attention, both theoretical and experimental [1-7].

Thus, from the standpoints of practical device application and fundamental science, the influence of quantum effects on the piezoresistance of two-dimensional heterostructures has been investigated [8]. Moreover, since cantilever based sensors have become a standard on micro- and nano-electromechanical systems (MEMS and NEMS) for detection of magnitudes with resolution in the pico-scale, there is recent work in developing electromechanical models for a transducer based on a lateral resonating cantilever [9] and a feedback controlled nanocantilever device [10]. Additionally, very recently, Qi Ye, et al, have beginning a large-scale fabrication of carbon nanotube probe tips for atomic force microscopy imaging applications [11].

Therefore, the study and understanding of electronic properties of systems in the nano-scale is very important for a wide range of possible applications. One example of this is the modeling of the tip and sample surfaces in scanning tunneling microscopy as confocal hyperboloids [12] and the consequent construction of their normal modes and Green functions by using prolate spheroidal coordinates [13]. According to this motivation and the experience obtained in the study of Stark effect with variational techniques in several geometries [7, 14], we decided to study this effect in the tip of a typical cantilever of an atomic force microscope (AFM) using as a first approximation the geometry given by a slice of a cake since, as a matter of fact, there exist commercial triangular-shaped silicon microcantilevers (Thermomicroscopes) [15]. Consequently, in the present work, we carry out a variational calculation to study the influence of an external applied electric field on the electronic ground-state energy of a slice of a cake as function of the parameters of the slice (diameter, angular aperture and thickness) and the intensity of the electric field, when this is applied along the axis of the slice directed to its wider part. For our calculations we suppose that the dielectric constant of the slice is the same as the medium where it is embedded.

## 2. Calculation

In cylindrical coordinates, the Hamiltonian of a carrier in a quantum box defined by the geometry of a slice of a cake with an electric field applied along its symmetry axis as shown in Fig. 1, is given by

$$H = \frac{p^2}{2m^*} + |e|F\rho\cos\theta + V_c(\rho,\theta,z), \tag{1}$$

where $F>0$ is the electric field, $m^*$ and $|e|$ are the electron effective mass and charge, respectively, and $V_c$ is the confining potential, which vanishes inside the slice and becomes infinite outside using the infinite well model.

For the ground-state of the system in the absence of the field, an exact calculation gives the corresponding wave function as

$$\Psi_0(\rho,\theta,z) = N_0 J_{m_0}\left(\frac{\alpha_{m_0,1}}{d}\rho\right)\cos\left(\frac{\pi}{\theta_0}\theta\right)\cos\left(\frac{\pi}{L}z\right), \tag{2}$$

with $N_0$ the normalization constant, $m_0 = \frac{\pi}{\theta_0}$ and $\alpha_{m_0,1}$ is the 1st zero of the Bessel function of order $m_0$, $J_{m_0}$. Evidently, $\Psi_0(\rho,\theta,z)=0$ outside the slice.

Now, when the electric field is applied, we will use as our variational function this ground-state wave function multiplied for an exponential factor, which accounts for the Stark effect, as follows:

$$\Psi(\rho,\theta,z) = \Psi_0(\rho,\theta,z)\exp(-\beta\rho\cos\theta), \tag{3}$$

with $\beta$ the variational parameter that depends on the electric field. Then, we define the Stark shift as

$$\Delta E = \langle\Psi|H|\Psi\rangle - \frac{\hbar^2}{2m^*}\left[\left(\frac{\alpha_{m_0,1}}{d}\right)^2 + \left(\frac{\pi}{L}\right)^2\right], \tag{4}$$

where the second term of the right side is the ground-state energy in absence of the electric field.

## 3. Results and discussion

All the results will be presented in reduced atomic units (a.u.*), which correspond to a length unit of an effective Bohr radius, $a^* = \hbar^2\varepsilon/m^*e^2$, and an effective Rydberg, $R^* = m^* e^4 / 2\hbar^2 \varepsilon^2$. The electric field is also given in atomic units as $F_0 = e / 2\varepsilon a^{*2}$. The dielectric constant, $\varepsilon$, is equal to 1 as it was stated above.

Obviously, the Stark shift does not depend on the thickness $L$ since the confinement due to the electric field is in the perpendicular direction as it is shown in the Hamiltonian, Eq. (1). Hence, for simplicity and in order to better simulate the tip of the cantilever, we chose for the rest of the calculations to take the thickness as $L/a^*=1$.

For an electric field applied in the positive direction of the $x$-axis (i.e. to the wider part of the slice), the electron is pushed to the tip of the slice. Figure 2 shows the Stark shift as a function of radius for angles $\theta < \pi$. In this figure, it can be observed that the Stark shift of the system increases as the electric field increases and it is larger for smaller angles. These two facts are a consequence of the quantum confinement of the electron in the tip since this confinement is responsible of an increment of the kinetic energy of the electron. It is also evident that the bigger the slice radio, the stronger the Stark shift since there is more radial space to confine the electron in the tip of the slice (i.e.) it is more notorious the confinement effect mentioned above.

In Fig. 3, $\Delta E$ is shown as a function of the electric field for various slice radii. This figure makes more evident this quantum confinement of the electron: in the case $\theta_0 = \pi/20$, it can be observed that for a given radius, the electric field enhances the energy of the electron since this is more strongly confined to the tip of the slice.

## 4. Conclusions

We have analyzed the effect of an applied electric field on the electronic ground-state energy of a quantum box with a geometry defined by a slice of a cake. We have shown how the quantum confinement effect strongly depends on the geometry of the system and the intensity of the electric field. When the field is applied directed to the wider part of the slice, there is a strong electronic confinement in the tip of the slice producing an increment of the electron energy. In order to make a more realistic modeling of the tip of a typical cantilever of an atomic force microscope, application of the electric field in the opposite direction, considering more complicated applied electric fields and a dielectric constant different to one as well as the inclusion of impurities in the system; will be matter of future works.

## Acknowledgements

This work has been partially supported by grant IN-106201 (DGAPA, UNAM, México). J. A. Reyes-Esqueda also acknowledges fruitful discussions with Lorea Chaos Cador.

**Figure Captions**

**Fig. 1.** The geometry of the slice of a cake and the electric field applied along it directed to the wider part, which corresponds to the positive direction of the axis *x*. $0 \leq \rho \leq d$; $-\theta_0/2 \leq \theta \leq \theta_0/2$; $-l/2 \leq z \leq l/2$.

**Fig. 2**. The Stark shift of the electron's energy, $\Delta E$, is shown as a function of the slice radius for two electric fields.

**Fig. 3**. $\Delta E$ is shown, for the same angular apertures as in figure 2, as a function of the applied electric field.

Figure 1.

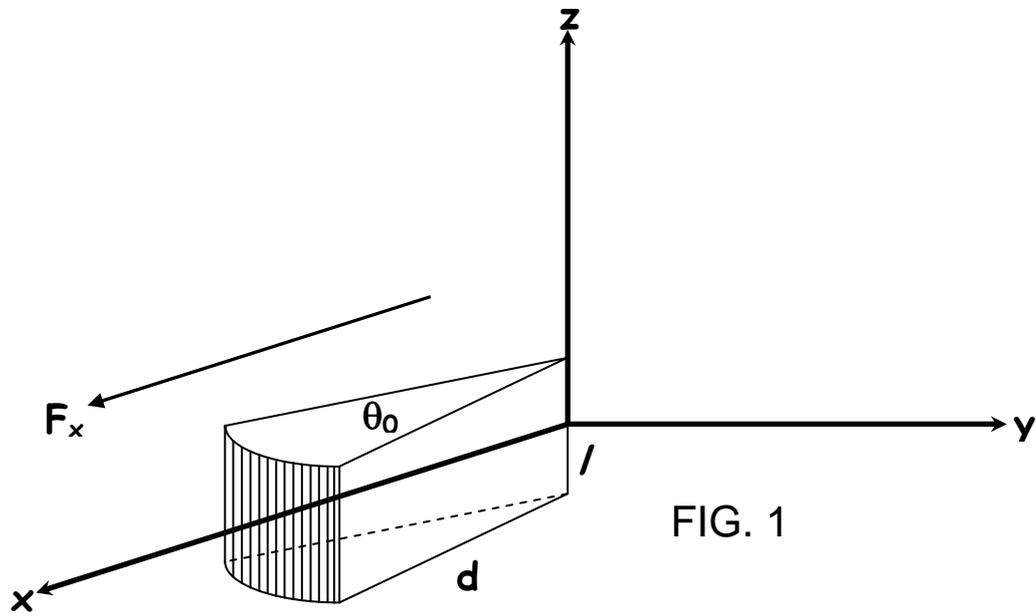

FIG. 1

*J. A. Reyes-Esqueda, et al.*

Figure 2.

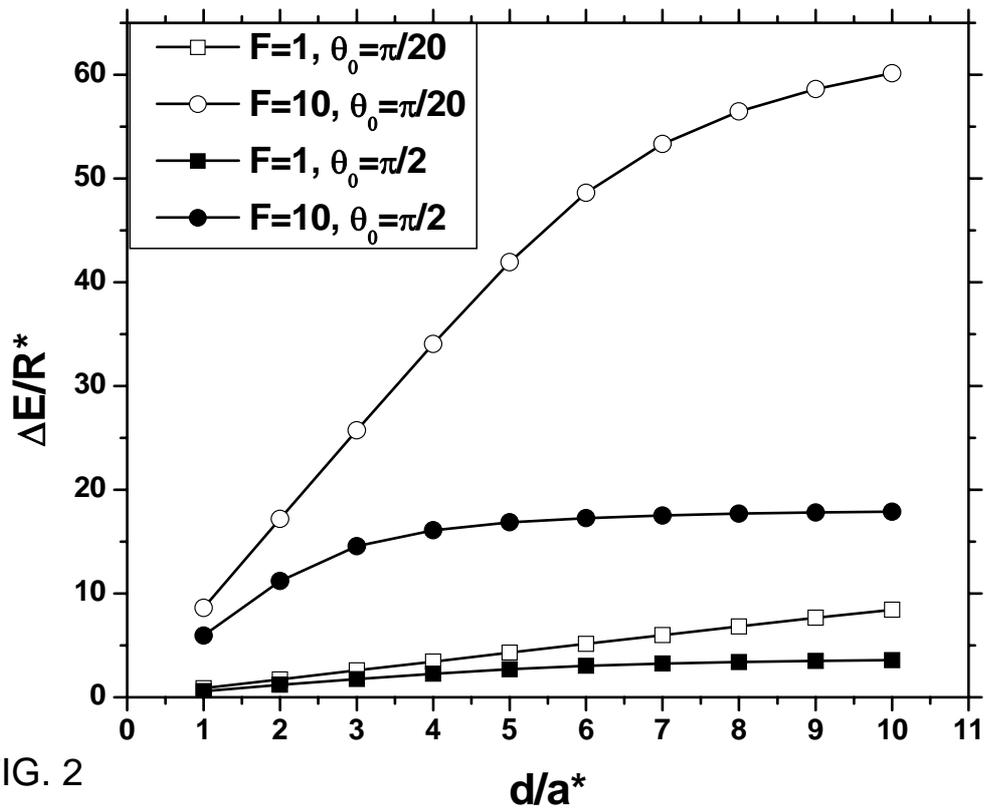

FIG. 2

*J. A. Reyes-Esqueda, et al.*

Figure 3.

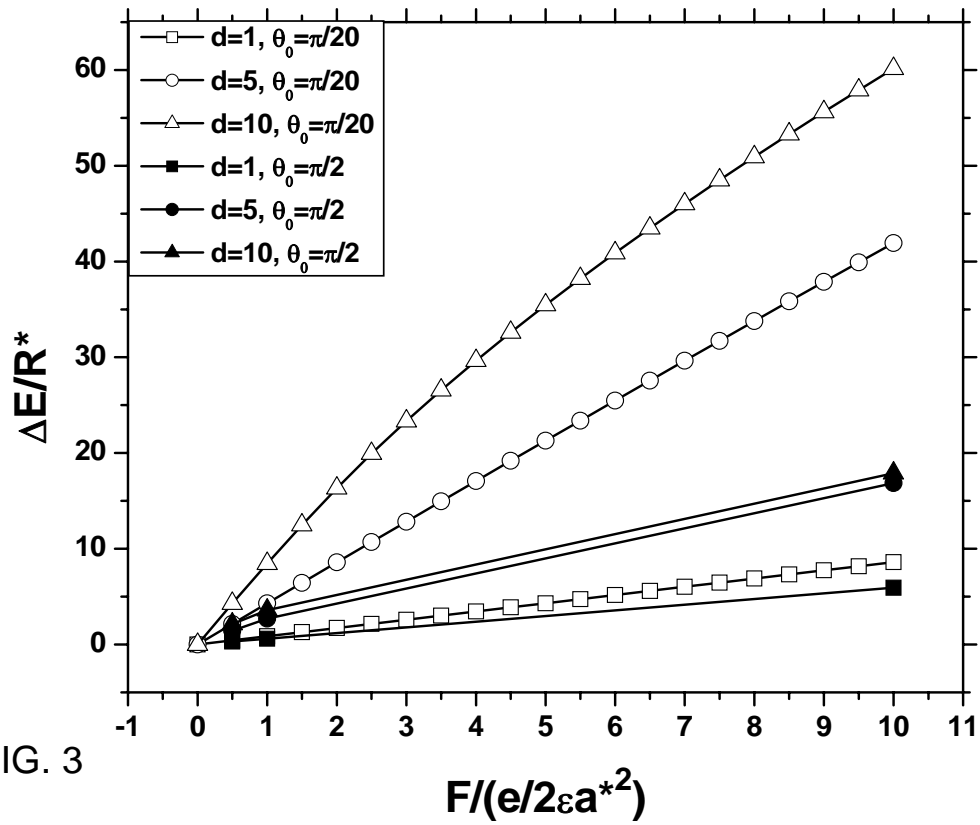

FIG. 3

*J. A. Reyes-Esqueda, et al.*